\documentclass[aps,twocolumn,pra]{revtex4-1}
\usepackage{dcolumn}
\usepackage{bm}
\usepackage{mathtools}
\usepackage{amsmath}
\usepackage{hyperref}
\hypersetup{colorlinks=true, urlcolor=blue}
\usepackage{color}
\usepackage{tikz}
\usepackage{multirow}
\usepackage{float}
\usepackage[caption = false]{subfig}
\usepackage{newlfont}
\usepackage{amssymb}
\usepackage{amsfonts}
\usepackage{amsthm}
\usepackage{graphicx}
\usepackage{bm}

\def \be{\begin{equation}}
\def \ee{ \end{equation} }

\begin{document}

\definecolor{red}{rgb}{1,0,0}
\title{Lower bounds on violation of monogamy inequality for quantum correlation measures}
\author{Asutosh Kumar}
\email{asukumar@hri.res.in}
\author{Himadri Shekhar Dhar}
\email{himadrisdhar@hri.res.in}

\affiliation{Harish-Chandra Research Institute, Chhatnag Road, Jhunsi, Allahabad 211 019, India}

\date{\today}

\begin{abstract}
In multiparty quantum systems, the monogamy inequality proposes an upper bound on the distribution of bipartite quantum correlation between a single party and each of the remaining parties in the system, in terms of the amount of quantum correlation shared by that party with the rest of the system taken as a whole.
However, it is well-known that not all quantum correlation measures universally satisfy the monogamy inequality. 
In this work, we aim at determining the non-trivial value by which the monogamy inequality can be violated by a quantum correlation measure. Using an information-theoretic complementarity relation between the normalized purity and quantum correlation in any given multiparty state, we obtain a non-trivial lower bound on the negative monogamy score for the quantum correlation measure. 
In particular, for the three-qubit states the lower bound is equal to the negative von Neumann entropy of the single qubit reduced density matrix.
We analytically examine the tightness of the derived lower bound for certain $n$-qubit quantum states. Further, we report numerical results of the same for monogamy violating correlation measures using Haar uniformly generated three-qubit states.
\end{abstract}

\maketitle

\section{Introduction}
\label{sec:intro}

The concept of monogamy \cite{mono1,mono2,mono3} is an intrinsic aspect in the study of quantum correlations in multiparty systems and plays a precursory role in quantum security \cite{mono3, qkd, nosignal} and multiport communication protocols such as super dense-coding \cite{nepal}. 
In its most primitive avatar, monogamy of quantum correlation states: \textit{if two parties \textit{A} and \textit{B} in a multiparty quantum state (say, $\varrho_{ABC}$) are maximally quantum correlated ($\mathcal{Q}(\varrho_{AB}) = 1$), then  neither \textit{A} nor \textit{B} can possess any quantum correlation with a third party, i.e., $\mathcal{Q}(\varrho_{AC}) = \mathcal{Q}(\varrho_{BC}) = 0$.} 
The above statement is satisfied by all quantum correlation measures, and is a clear departure from classical correlations which are not bound to such constraints. 
Interestingly, monogamy property is not limited to quantum correlations but is manifested in other quantum properties such as Bell inequality violations \cite{bell}, quantum steering \cite{steer}, channel capacities \cite{densehri}, and contextual inequalities \cite{context}.

However, in general, two parties in a quantum state need not necessarily share maximal quantum correlation, and are thus able to share some correlations with other parties, albeit in a constrained way. The concept of monogamy thus restricts the distribution of bipartite quantum correlations in multiparty quantum systems but its quantification is not always achievable.
%
For three-qubit systems, an important figure of merit, the \textit{monogamy inequality}, was obtained for an entanglement measure called the tangle \cite{monogamy}. It was shown that, for a general three-qubit state, $\varrho_{ABC}$, the amount of tangle ($\tau$) shared between the party $A$, individually with parties $B$ and $C$, is bounded above by the tangle shared by $A$ with $BC$ taken as a whole. Mathematically, this implies: 
$\tau(\varrho_{A:BC}) \geq \tau(\varrho_{A:B})$ + $\tau(\varrho_{A:C})$. It is easy to verify, that the inequality satisfies the aforementioned statement for monogamy. If $\tau(\varrho_{A:B})$ = 1, then $\tau(\varrho_{A:C})$ is necessarily zero, as $\tau(\varrho_{A:BC}) \leq$ 1. Moreover, the monogamy inequality for tangle was shown to exist even in $n$-qubit quantum states \cite{monogamy2}. In literature, states and quantum correlations that satisfy the above monogamy inequality are termed as \textit{monogamous}. However, it is well known that the monogamy inequality is not satisfied universally for all quantum correlation measures, even for three-qubit states \cite{asu-uni, giorgi11-pra84, prabhu12-pra85r,  fanchini11-pra84, bruss12-prl109, fanchini13-pra87}. Entanglement measures such as entanglement of formation \cite{eof}, apart from information-theoretic measures such as quantum discord \cite{discord1} and known to be, in general, non-monogamous. 
Recent results on monogamy have shown that quantum correlation measures may satisfy the monogamy inequality by minute extensions of the upper bound \cite{fanchini14-pra89}, increasing fractional and integer exponents of quantum correlation measures \cite{salini14-aop348}, or by considering large number of parties \cite{asu-multi} (cf. \cite{comm}).

For all quantum correlation measures, one can define a \textit{monogamy score} \cite{manab12-pra86}, which captures the difference between the non-negative quantities, $\mathcal{Q}(\varrho_{A:\mathrm{rest}})$, and $\sum_k \mathcal{Q}(\varrho_{A:B_k})$, in a quantum state, $\varrho_{A B_1 B_2 \cdots B_n}$. The monogamy score ($\delta_\mathcal{Q}$) is postive for all measures that satisfy the monogamy inequality, and is bounded above by the term $\mathcal{Q}(\varrho_{A:\mathrm{rest}})$. In recent works, for $n$-qubit pure quantum states, the upper bound for all quantum measures, regardless of the fact that they satisfy the inequality, has been expressed as functions of the genuine multipartite entanglement  \cite{prabhu12-pra86,asu-forb}.
The lower bound for $\delta_\mathcal{Q}$ is trivially zero for situations where the inequality is satisfied (as $\delta_\mathcal{Q} \geq$ 0). However, for quantum correlation measures that do not satisfy the monogamy inequality, there does not exist a non-trivial lower bound. It is then natural to ask whether there exists a bound on the value by which the monogamy inequality can be violated. 
{The existence of a non-trivial lower bound on $\delta_\mathcal{Q}$ is an important aspect in the study of various quantum information protocols. Monogamy poses a fundamental restriction on the sharability and distribution of quantum resources, and while this provides a significant advantage in obtaining bounds on secret key rates in quantum cryptography \cite{qkd}, it also restricts the availability of resources in 
multipartite protocols such as dense coding \cite{densehri}. 
Moreover, the monogamy score has been used as an important figure of merit in studies on multiparty quantum states \cite{prabhu12-pra85r, manab12-pra86}, identification of quantum channels \cite{asu-new}, and distinguishing phases of many-body systems \cite{aditi-nmr}. 
However, the monogamy score is a difficult quantity to compute and estimate for generic quantum states and generic quantum correlations. It is therefore interesting to derive both upper and lower bounds of the monogamy score. While the upper bounds of monogamy score are known in the literature  \cite{prabhu12-pra86,asu-forb}, 
our study aims at obtaining a non-trivial lower bound of the quantity.}

In this work, we determine a non-trivial lower bound on the negative monogamy score of quantum correlation measures that violate the monogamy inequality, using an information-theoretic complementarity between the normalized purity of the reduced state ($\varrho_{A}$) and quantum correlation shared between two parties ($\varrho_{AB}$). We observe that, under certain conditions, for three-qubit quantum states the lower bound reduces to the  negative of the von Neumann entropy \cite{vonNeumannentropy} of the single-qubit reduced density matrix. Using a set of well-known quantum states, we analytically derive the lower bound for larger number of parties and examine its variation and tightness. Moreover, we numerically evaluate the bound for a set of entanglement and information-theoretic quantum correlation measures by randomly generating Haar-uniform three-qubit states, to show that the bound is satisfied.
This paper is organized as follows.
In Sec.~\ref{sec:def}, we briefly review the monogamy of quantum correlation measures, and the complementarity relation between the normalized purity of a subsystem with respect to the bipartite quantum correlation ${\cal Q}$ in the quantum system. Lower bounds on monogamy scores, in terms of the complementarity relation, are determined in Sec.~\ref{sec:lowerbound}. We then analyze, in Sec.~\ref{anal}, the characteristics of the lower bound using a few well-known quantum states and quantum correlation measures. 
We conclude in Sec.~\ref{sec:conclude}.%

\section{Monogamy and complementarity}
\label{sec:def}
We begin with a brief discussion on the monogamy inequality, as introduced by Coffman \emph{et al.} for the tangle of three-qubit states \cite{monogamy}, and later expanded for higher number of qubits \cite{monogamy2} and other quantum correlation measures.
%
%
%
%
Consider that ${\cal Q}$ is a bipartite quantum correlation measure. For a multipartite quantum system described by state $\varrho_{AB_1 B_2\ldots B_n} \equiv \varrho_{AB}$, the monogamy property is captured by the inequality 
\begin{equation}
\label{eq:std-mono-ineq}
\sum_{j=1}^n{\cal Q}(\varrho_{AB_j}) \leq {\cal Q}(\varrho_{AB}),
\end{equation}
where the party $A$ is the nodal observer.
Quantum correlation measures that universally satisfy the above inequality is said to satisfy monogamy or is {monogamous}. 
Similarly, the quantum state, $\varrho_{AB_1 B_2\ldots B_n}$, is said to be monogamous under the quantum correlation measure ${\cal Q}$, if it satisfies Eq.~(\ref{eq:std-mono-ineq}).
We note that the state need not be monogamous for all quantum correlations measures, and in general, can be non-monogamous. 
An important figure of merit to investigate the monogamy inequality is the monogamy score \cite{manab12-pra86}. It encapsulates the deficit between the quantities in the monogamy inequality, and is defined as
\begin{equation}
\delta_{{\cal Q}}={\cal Q}(\varrho_{AB}) - \sum_{j=1}^n{\cal Q}(\varrho_{AB_j}).
\label{eq:mono-score}
\end{equation}
An obvious extension of the monogamy inequality is that the $\delta_{{\cal Q}}$ is non-negative ($\delta_{{\cal Q}} \geq 0$) for all monogamous $\cal Q$. One can interpret a positive $\delta_{{\cal Q}}$ as ``residual'' quantum correlation of an \(n\)-party quantum state that is not captured by the distributed bipartite quantum correlations, which in essence captures some form of multipartite correlations in the system \cite{prabhu12-pra86}. Several quantum correlation measures, including entanglement monotones such as tangle \cite{monogamy}, squashed entanglement \cite{koashi} and squared negativity \cite{sq-neg}, are monogamous, as opposed to measures such as entanglement of formation \cite{eof}, quantum discord \cite{discord1}, negativity \cite{negativity}, logarithmic negativity \cite{logneg}, and quantum work-deficit \cite{qwd,qwd1}.



We now discuss an information-theoretic complementarity relation, for multipaty quantum states, between the normalized purity of a subsystem of the state and the bipartite quantum correlation the subsystem shares with rest of the system, as derived in Ref.~\cite{asu-comp}. 
Let us consider a bipartite quantum correlation measure ${\cal Q}$, for quantum state $\varrho_{XY} \in \mathbb{C}^{d_X} \otimes \mathbb{C}^{d_Y}$, where $X$ and $Y$ are two subsystems of the state. Let us say, ${\cal Q}$ satisfies the conditions (i) ${\cal Q}(\varrho_{XY}) \leq S(\varrho_X)$, which is true for a host of quantum correlation measures \cite{horodecki09}, and (ii) $0 \leq {\cal Q}(\varrho_{XY}) \leq \log_2 d_Y$. For such measures, it has been shown that ${\cal Q}$ obeys a non-trivial complementarity relation between purity of the reduced state, $\varrho_{X}$, and the shared bipartite quantum correlation in the state, $\varrho_{XY}$, as given below \cite{asu-comp}:
\begin{equation}
\label{eq:cr}
{
 \mathbb{P}(\varrho_{X}) + \mathbb{Q}(\varrho_{XY}) \leq b \left.\mathrel{}\middle|\mathrel{}\right.
        \begin{cases}
        b = 1,  & \text{if } d_{X} \leq d_{Y}, \\
        b = 2 - \frac{\log_2 d_Y}{\log_2 d_{X}}, & \text{if } d_{X} > d_{Y},
        \end{cases}}
\end{equation}
where $\mathbb{P}$ is defined as,
\begin{equation}
\label{eq:purity}
\mathbb{P}(\varrho_{X})=\frac{\log_2 d_{X}-S(\varrho_{X})}{\log_2 d_{X}},
\end{equation}
and quantifies the normalized purity of the system \cite{note1} in the \(X\)-part, and $\mathbb{Q}(\varrho_{XY})$, given by
\begin{equation}
\label{eq:correl}
\mathbb{Q}(\varrho_{XY})=\frac{{\cal Q}(\varrho_{XY})}{\min \{\log_2 d_{X}, \log_2 d_{Y}\}},
\end{equation}
represents the normalized quantum correlation of the system in the $X$:$Y$ bipartition. 
For example, an important quantity that satisfies the complementarity relation is the quantum mutual information between two parties, defined as, ${\cal I}(\varrho_{XY})$ = $S(\varrho_X)$ + $S(\varrho_Y)$ - $S(\varrho_{XY})$ \cite{qmi}.  ${\cal I}(\varrho_{XY})$, which captures the total correlations in a system, $\varrho_{XY}$, satisfies the condition
${\cal I}(\varrho_{XY}) \leq 2\min \{S(\varrho_X),S(\varrho_Y)\} \leq 2\min \{\log_2 d_X, \log_2 d_Y\}$.
Hence if one considers the correlation, ${\mathcal Q}(\varrho_{XY})$ = ${\cal I}(\varrho_{XY})/2$, the complementarity is naturally satisfied.
Moreover, for tripartite quantum states, $\varrho_{ABC} \in \mathbb{C}^d \otimes \mathbb{C}^d \otimes \mathbb{C}^d$, the complementarity relation reduces to 
\begin{equation}
 \mathbb{P}(\varrho_{AB}) + \mathbb{Q}(\varrho_{AB:C}) \leq \frac32,
\end{equation} 
and is independent of the dimension of the system, and can be shown to be saturated by the Greenberger-Horne-Zeilinger (GHZ) state \cite{ghz}.

This complementarity relation has potential application in quantum information protocol. In particular, it has been used to obtain a security proof of quantum cryptography for individual attacks via a variation \cite{acin06-njp8} of the Ekert key distribution \cite{mono1} protocol. 
We note that the complementarity relation in Eq.~(\ref{eq:cr}), is satisfied by a host of quantum correlation measures, irrespective of the conditions (i) and (ii), as shown in Appendix \ref{app} (cf. \cite{asu-comp}). 

\begin{figure*}[]
\begin{center}
\subfloat[]{\includegraphics[width=5cm]{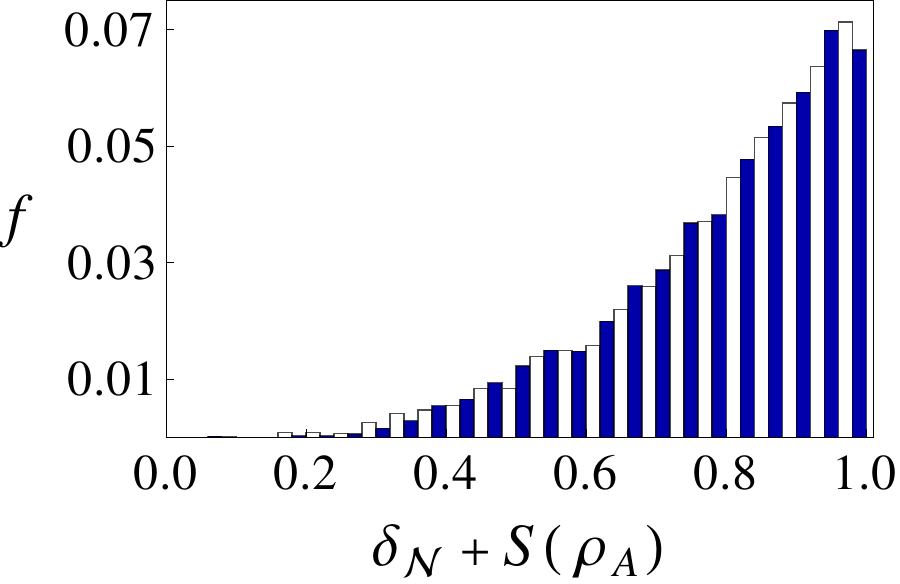}}\hspace{.4cm}
\subfloat[]{\includegraphics[width=5cm]{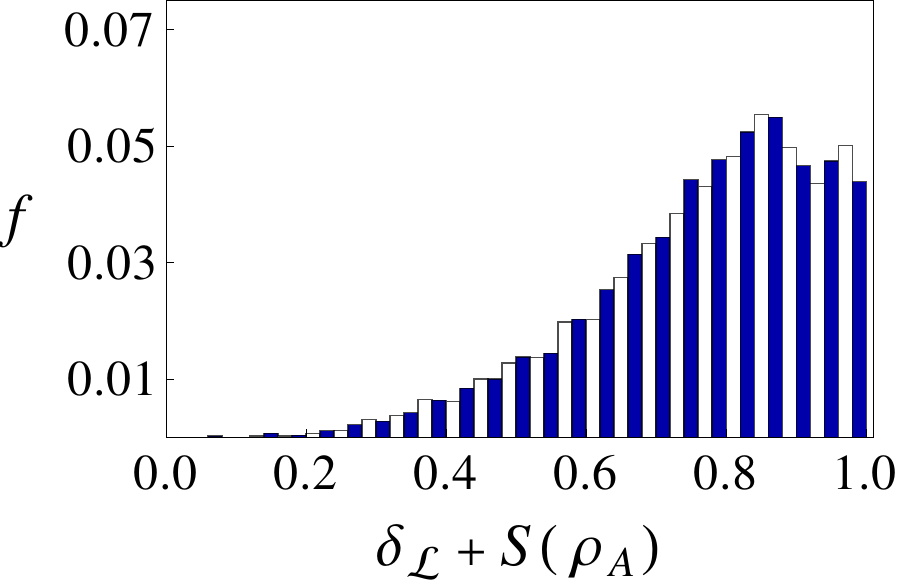}}\hspace{.4cm}
\subfloat[]{\includegraphics[width=5cm]{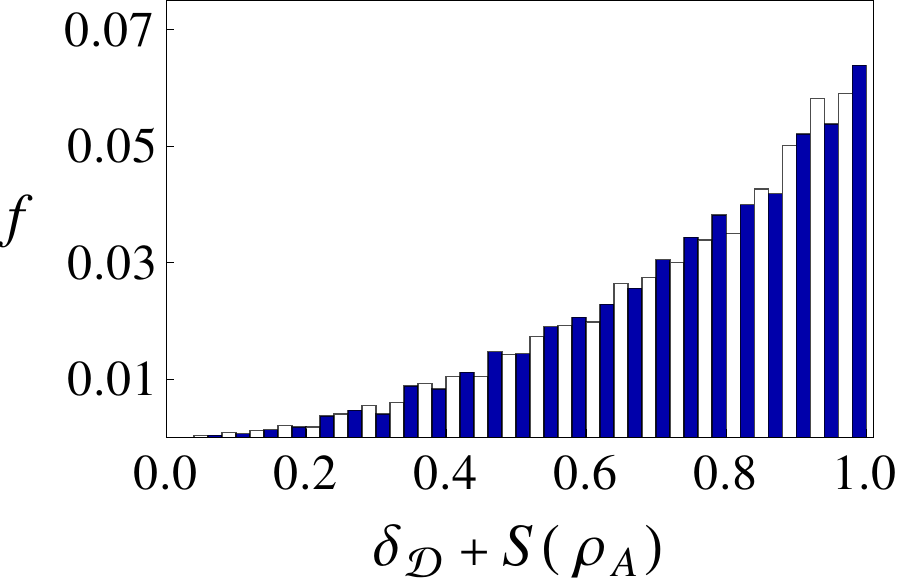}}\\
\subfloat[]{\includegraphics[width=5cm]{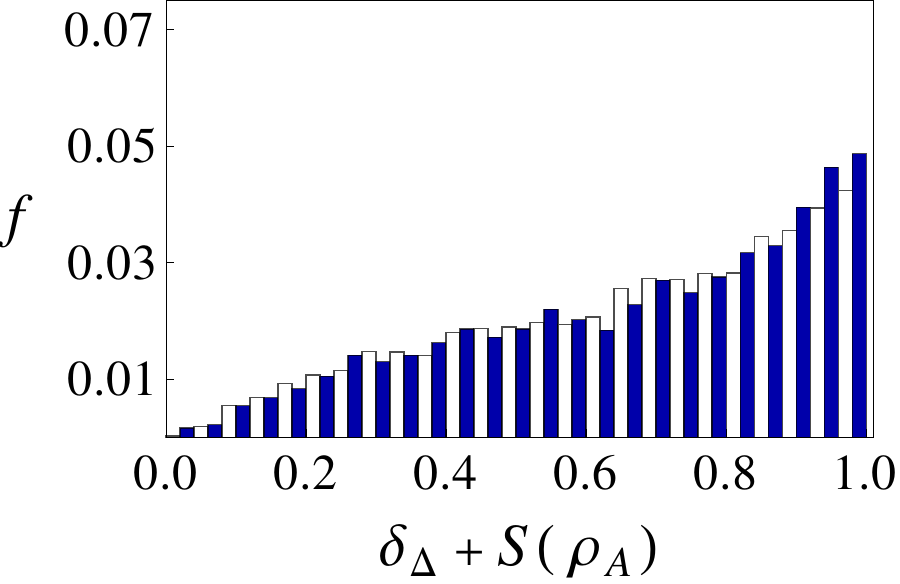}}\hspace{.4cm}
\subfloat[]{\includegraphics[width=5cm]{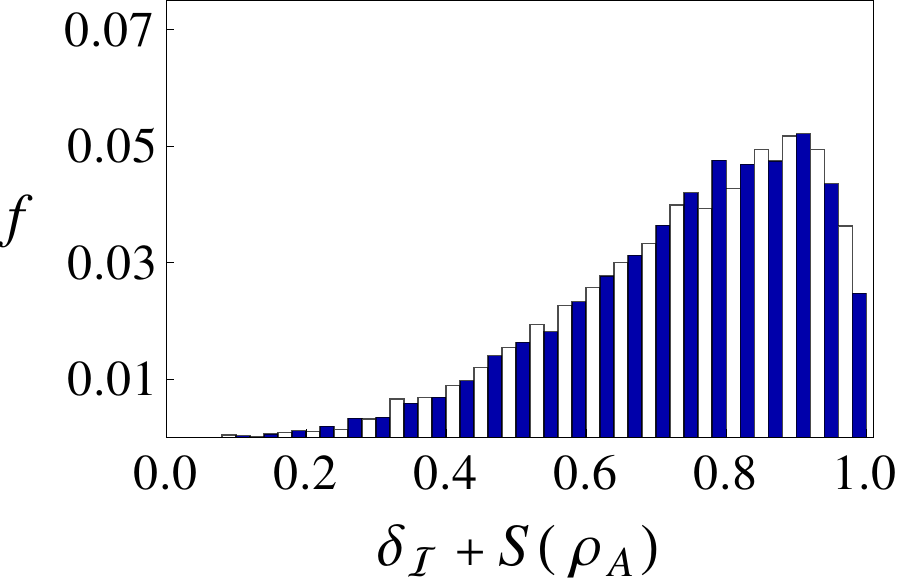}}\hspace{.4cm}
\subfloat[]{\includegraphics[width=5cm]{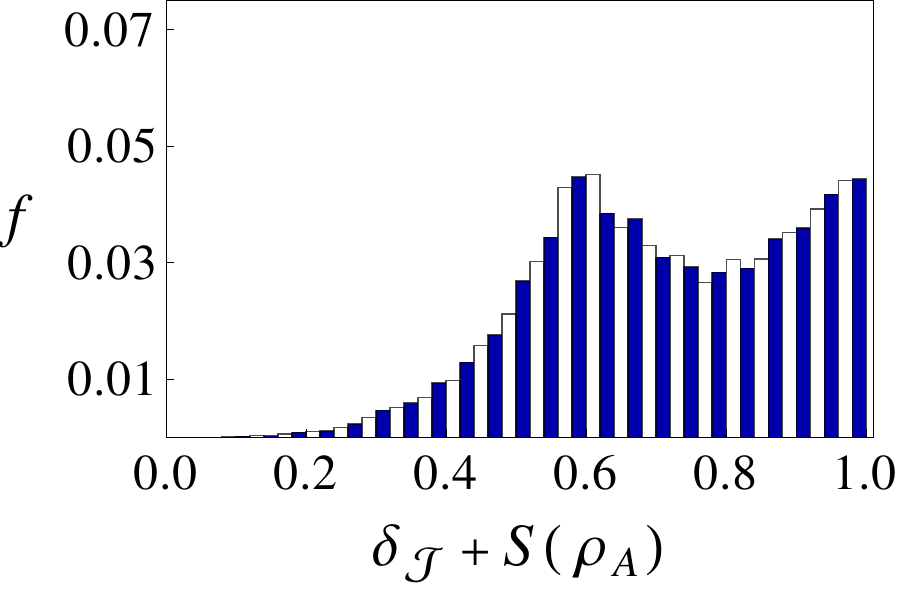}}
\caption{Histograms depicting the frequency ($f$) of states for the quantity $\delta_{\mathbb{Q}}$  + $S(\varrho_A)$, for different quantum correlation measures, $\mathbb{Q}$.  
The graph corresponds to a set ($\sim 10^6$) of random Haar uniform generated rank-1 and rank-2 three-qubit states. The monogamy score is calculated for the quantum correlation measures given by (a) negativity ($\mathcal{N}$), (b) logarithmic negativity ($\mathcal{L}$), (c) quantum discord ($\mathcal{D}$), (d) quantum work-deficit ($\Delta$), (e) quantum mutual information ($\mathcal{I}$), and (f) conditional mutual information ($\mathcal{J}$). The histogram shows that $\delta_{\mathbb{Q}}$  + $S(\varrho_A)$ $\geq$ 0, and thus satisfies 
$\delta_{\mathbb{Q}} \geq -S(\varrho_A)$. }
\label{fig1}
\end{center}
\end{figure*}

\section{Lower bounds on monogamy scores}
\label{sec:lowerbound}

An important aspect of the monogamy score, in Eq.~(\ref{eq:mono-score}), is that the quantity has a distinct upper bound given by the term ${\cal Q}(\varrho_{AB})$, where we use $\varrho_{AB}$ = $\varrho_{A(B_1\cdots B_n)}$. This implies that for monogamous ${\cal Q}$, the amount of distributed bipartite entanglement in a multipartite state is bounded by the amount of block entanglement shared by a single party with the rest of the system. Interestingly, it has been shown that the upper bound of $\delta_{{\cal Q}}$, for a large number of pure $n$-qubit states, is determined by a set of entropic or quadratic functions of the genuine multipartite entanglement \cite{ggm,biswas14-pra90} of the state $\varrho_{AB}$ \cite{prabhu12-pra86,asu-forb}.
Moreover, for monogamous ${\cal Q}$, there exists a definite lower bound on $\delta_{{\cal Q}}$, given by $\delta_{{\cal Q}} >$ 0. However, for ${\cal Q}$ that does not satisfy monogamy inequality, in general, the situation is not straightforward. Though the upper bound remains the same, a non-trivial lower bound is not obvious. 
For cases where $\varrho_{AB}$ is not monogamous with respect to ${\cal Q}$, $\delta_{{\cal Q}}$ can be negative. One of the primary motivation of our work is to find out the non-trivial degree or limit to which this violation of monogamy inequality occurs. In other words, we question, \textit{what is the worst negative value of the quantity $\delta_{{\cal Q}}$?}

In this section, we aim at obtaining the lower bound on monogamy scores, for non-monogamous measures ${\cal Q}$, by making use of the complementarity relation, given in Eq.~(\ref{eq:cr}). 
In deriving the lower bounds on $\delta_{{\cal Q}}$ below,
we will use the notation ${\cal Q}_{XY} \equiv {\cal Q}(\varrho_{XY})$, $\mathbb{Q}_{XY} \equiv \mathbb{Q}(\varrho_{XY})$, $\mathbb{P}_X \equiv \mathbb{P}(\varrho_X)$, among others, and where 
\(\varrho_{X} = \mbox{Tr}_Y \varrho_{XY}\). Consider an arbitrary multipartite quantum state $\varrho_{AB} \equiv \varrho_{A(B_1B_2\ldots B_n)}$ of $n+1$ parties. For a normalized bipartite quantum correlation measure $\mathbb{Q}$, a trivial lower bound on the monogamy score is provided in terms of the number of parties in the system $B$ (= $B_1B_2\ldots B_n$), as given by 
\begin{equation}
\label{eq:lbound1}
\delta_{\mathbb{Q}} = \mathbb{Q}_{AB} - \sum_{k=1}^n \mathbb{Q}_{AB_k} \geq -(n-1).
\end{equation} 
To obtain a non-trivial lower bound on the $\delta_{{\mathbb Q}}$ for non-monogamous measures, we use the complementarity relation as shown below. 
For $\rho_{AB}$, using the relation in Eq.~(\ref{eq:cr}), we can write 
\begin{align}
\label{eq:lbound2}
\mathbb{P}_{A} + \mathbb{Q}_{AB} &= x_0~ (\mathrm{say})~ \leq b_0 \nonumber \\
\mathit{or,}~~ \mathbb{Q}_{AB} &= x_0 - \mathbb{P}_{A}.
\end{align}

Similarly, for the reduced density matrices, $\varrho_{AB_k}$, where all parties except $A$ and $B_k$ have been traced out, i.e., $\varrho_{AB_k}$ = $\mbox{Tr}_{\overline{AB_k}} (\varrho_{AB})$, we obtain $\mathbb{P}_{A}$ + $\mathbb{Q}_{AB_k}$ = $x_k \leq b_k$. Summing over all $k$, we get the relation
\begin{equation}
\label{eq:lbound3a}
n \mathbb{P}_{A} + \sum_{k=1}^n \mathbb{Q}_{AB_k} = \sum_{k=1}^n x_k \leq \sum_{k=1}^n b_k.
\end{equation}
When $d_A \leq d_{B_k}$ (say, $d_A = d_{B_k} = d$), using the complementarity relation (\ref{eq:cr}), one has $b_k$ = 1, $\forall~b_k$. Hence, Eq.~(\ref{eq:lbound3a}) can be written as
\begin{align}
\label{eq:lbound3b}
n \mathbb{P}_{A} + \sum_{k=1}^n \mathbb{Q}_{AB_k} &= \sum_{k=1}^n x_k \leq n \nonumber \\
\mathrm{or,}~~ \sum_{k=1}^n \mathbb{Q}_{AB_k} &\leq n(1 - \mathbb{P}_{A}).
\end{align}
Subtracting Eq.~(\ref{eq:lbound3b}) from Eq.~(\ref{eq:lbound2}), we obtain an expression for the monogamy score, $\delta_{\mathbb{Q}}$, which upon arranging, gives us
\begin{equation}
\label{eq:lbound4a}
\delta_{\mathbb{Q}} = \mathbb{Q}_{AB} - \sum_{k=1}^n \mathbb{Q}_{AB_k} \geq -(n-1)(1 - \mathbb{P}_{A}) - (1- x_0).
\end{equation}
By rearranging Eq.~(\ref{eq:lbound4a}), one can show that the obtained lower bound on monogamy score can be improved, in comparison to Eq.~(\ref{eq:lbound1}), as is evident from
\begin{equation}
\label{eq:lbound4c}
\delta_{\mathbb{Q}} \geq -(n-1)\left(1 - \mathbb{P}_{A} + \frac{1- x_0}{n-1} \right).
\end{equation}
For large $n$ or for $x_0 \geq 1$ \cite{comment}, the lower bound on monogamy score reduces to 
\begin{equation}
\label{eq:lbound5}
\delta_{\mathbb{Q}} \geq -(n-1)(1 - \mathbb{P}_{A}),
\end{equation}
where $\mathbb{P}_{A}$ is the normalized purity defined in Eq.~(\ref{eq:purity}), and ranges from $0 \leq \mathbb{P}_{A} \leq 1$.
For the important three-qubit states, which have received a lot of attention in the study of monogamy, we observe that the lower bound on the monogamy score is given by, $\delta_{\mathbb{Q}} \geq -S(\varrho_{A})$, where $S(\varrho_{A})$ is the von Neumann entropy of the subsystem $A$. This shows that the lower bound satisfies the limits for monogamy ($\delta_{\mathbb{Q}} \gtrsim$ 0) for states that are weakly entangled along the $A:rest$ bipartition, such that $S(\varrho_{A}) \approx$ 0.

\section{\label{anal}Analyzing the lower bound}

Let us now examine the lower bound of the monogamy score, $\delta_{\mathbb{Q}}$, for certain quantum states. For example, we consider $(n+1)$-party quantum states  $\varrho_{AB_1\cdots B_n} \in \mathbb{C}^2 \otimes \mathbb{C}^{d_{B_1}} \otimes \cdots \otimes \mathbb{C}^{d_{B_n}}$.
For these states, the lower bound in relation (\ref{eq:lbound5}) becomes $\delta_{\mathbb{Q}} \geq -(n-1)S(\varrho_A)$. For $d_{B_k}$ = 2, $\forall~ k$, the above expression for the lower bound of $\delta_{\mathbb{Q}}$ holds for all multiqubit quantum states.  

Let us consider the  superposition of the generalized GHZ \cite{ghz} and the \textit{W} state \cite{W,threeslocc}, which are permutationally invariant multiqubit states given by
\begin{equation}
|\Psi_{\alpha,\gamma}\rangle = \alpha|0\rangle^{\otimes n} + \beta|1\rangle^{\otimes n} + \gamma|W^{n}\rangle,
\end{equation}
where {$|W^{n}\rangle$ is the normalized $n$-qubit \textit{W} state, given by 
$|W^{n}\rangle$ = $\frac{1}{\sqrt{n}} \sum \mathcal{P}\left(|0\rangle^{\otimes(n-1)} \otimes |1\rangle\right)$, where the sum in the expression is over all $\cal{P}$ permutations of the product state containing $(n-1)$ $|0\rangle$'s, and a single $|1\rangle$.
$\alpha$, $\beta$, and $\gamma$ are complex numbers that satisfy the normalization constraint
$|\alpha|^2$ + $|\beta|^2$ + $|\gamma|^2$ = 1.} 
The $n$-qubit state, $|\Psi_{\alpha,\gamma}\rangle$, satisfies the lower bound for the monogamy score, given by $\delta_{\mathbb{Q}} \geq -(n-2)S(\varrho_{\alpha,\gamma})$, where $\varrho_{\alpha,\gamma}$ = $\mbox{Tr}_{n-1} (|\Psi_{\alpha,\gamma}\rangle\langle\Psi_{\alpha,\gamma}|$) is obtained by tracing out $n-1$ qubits.  $\varrho_{\alpha,\gamma}$ can be written as
%
%
\begin{equation}
\varrho_{\alpha,\gamma} = 
\left(\begin{array}{ccc}
|\alpha|^2+\frac{n-1}{n}|\gamma|^2 & \alpha\gamma^*\sqrt{\frac{1}{n}} \\
\alpha^*\gamma\sqrt{\frac{1}{n}} & \frac{1}{n}|\gamma|^2 + |\beta|^2
\end{array}\right).
\label{ghz-w}
\end{equation}
The largest eigenvalue of the single qubit state $\varrho_{\alpha,\gamma}$ is given by
\[
e = \frac{1}{2}\left(1+\sqrt{1-4|\alpha|^2|\beta|^2 - \frac{4(n-1)}{n}|\gamma|^2(|\beta|^2 + \frac{|\gamma|^2}{n})}\right).
\]
Hence, $S(\varrho_{\alpha,\gamma})$ is equal to $h(e)$, where $h(x)$ is the Shannon (binary) entropy \cite{Shannonentropy} of the variable $x$. The lower bound of the monogamy score is given by, $\delta_{\mathbb{Q}} \geq -(n-2)h(e)$. Hence, a tight lower bound on the monogamy score is obtained for low values of $h(e)$.

If we set, $\gamma$ = 0, then we obtain the generalized GHZ state, $|\Psi_{GHZ}\rangle = \alpha|0\rangle^{\otimes n} + \beta|1\rangle^{\otimes n}$,
and the largest eigenvalue of the single qubit reduced state is $\frac{1}{2}\left(1+\sqrt{1-4|\alpha|^2|\beta|^2}\right)$, which is independent of $n$. For states with $|\alpha|^2$ = $|\beta|^2$ = 1/2, $e$ = 1/2, and $h(e)$ = 1, which gives us a weak bound. However, for states with $|\alpha|^2|\beta|^2 \approx 0$, $e \approx 1$, which implies $h(e) \approx$ 0. For these states, the lower bound of the monogamy score is given by, $\delta_{\mathbb{Q}} \geq -\epsilon$, where $\epsilon\rightarrow$ 0. Alternately, if one sets, $\alpha$ = $\beta$ = 0, such that $\gamma$ = 1, one obtains the $n$-qubit $W$ state, given by 
$|\Psi_W\rangle$ = $|W^n\rangle$.
The maximum eigenvalue of the reduced state is then given by, $e$ = $\frac{n-1}{n}$, and $h(e)$ = $h(1/n)$. Hence, as $n$ increases, the quantity $h(e)\rightarrow$ 0, and hence we obtain, $\delta_{\mathbb{Q}} \gtrsim 0$.

To expand the study on $W$ states, we consider the $n$-qubit Dicke state \cite{dicke} containing all $\cal{P}$ permutations of $n-r$ qubits in $|0\rangle$ and $r$ qubits in $|1\rangle$ state, as given in the equation
\begin{equation}
|\Psi_{r,n}\rangle = \dbinom{n}{r}^{-1/2} \sum \mathcal{P}\left(|0\rangle^{\otimes(n-r)} \otimes |1\rangle^{\otimes r}\right).
\label{dick}
\end{equation}
For $r$ = 1, the Dicke state is the same as an $n$-qubit $W$ state, $|\Psi_W\rangle$, mentioned earlier. The single qubit reduced state eigenvalue, $e$, is equal to $\frac{r}{n}$ (or $\frac{n-r}{n}$), and hence $h(e)$ = $h(r/n)$. Hence, as the $\frac{r}{n}$ ratio decreases, $h(e)$ decreases, and we obtain, $\delta_{\mathbb{Q}} \gtrsim 0$. However, for $r$ = $n/2$, $h(e)$ = 1, and the lower bound on the monogamy score is weak.

\begin{figure}[t!]
\begin{center}
\includegraphics[width=8cm, angle=0]{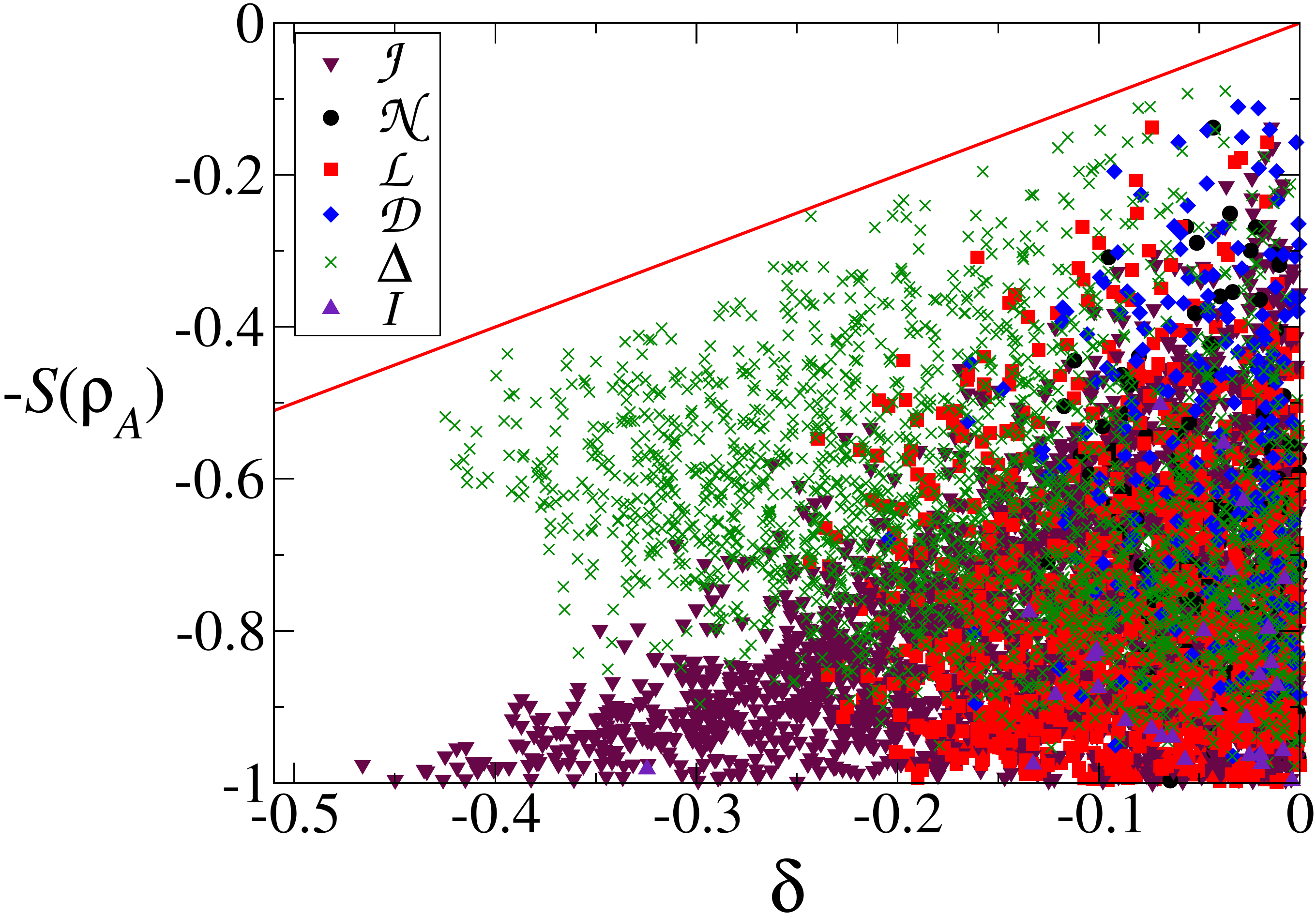}
\caption{(Color online.) Tightness of the bound $\delta_{\mathbb{Q}} \geq -S(\varrho_A)$.  
The score, $\delta \equiv \delta_{\mathbb{Q}}$, of monogamy inequality violating quantum correlation measures for a set of randomly generated Haar uniform three-qubit states, plotted along the $x$-axis, with respect to the negative von Neumann entropy ($-S(\varrho_A)$) of party $A$,  plotted along the $y$-axis. The measures are negativity (black-circle), logarithmic negativity (red-square), quantum discord (blue-diamond), quantum work-deficit (green-cross), quantum mutual information (indigo-up-triangle), and conditional mutual information (violet-down-triangle). 
The figure shows that $\delta_{\mathbb{Q}} \geq -S(\varrho_A)$. It is evident that the lower bound is tight for states with low reduced entropy. For the ease of viewing and without affecting the results, the plot is provided for a set of $2\times10^4$ rank-1 and rank-2 three-qubit states, drawn from a larger sample set ($\sim 10^6$) of randomly generated Haar uniform states. Along with Fig.~\ref{fig1}, the figure numerically reasserts the lower bound on monogamy score, given by 
$\delta_{\mathbb{Q}} \geq -S(\varrho_A)$.
}
\label{threequbit}
\end{center}
\end{figure}

Now we look at the monogamy score for a set of quantum correlation measures. We generate a large number of random Haar uniform three-qubit states, and compute $\delta_{\mathbb{Q}}$ for a set of monogamy inequality violating measures such as negativity ($\mathcal{N}$), logarithmic negativity ($\mathcal{L}$), quantum discord ($\mathcal{D}$), quantum work-deficit ($\Delta$), quantum mutual information ($\mathcal{I}$), and conditional mutual information ($\mathcal{J}$). For the generated pure and mixed three-qubit states, and for the above measures, it is observed that the monogamy score satisfies the lower bound, $\delta_{\mathbb{Q}} \geq -S(\varrho_A)$, as shown in Fig.~\ref{fig1}, which shows an histogram of the frequency of states against the quantity 
$\delta_{\mathbb{Q}}$ + $S(\varrho_A)$. The figure shows that $\delta_{\mathbb{Q}}$ + $S(\varrho_A)$ is always positive. To check the tightness of the bound, we plot the monogamy score, $\delta_{\mathbb{Q}}$, against $-S(\varrho_A)$ in Fig.~\ref{threequbit}. It is evident from the figure that the lower bound is tight for states with low reduced von-Neumann entropy.

\section{\label{sec:conclude} Conclusion}
Quantum correlations are an important aspect of modern physics and a key enabler in quantum communication and computation technologies. The concept of monogamy is a distinguishing feature of quantum correlations, which sets it apart from classical correlations, and has played a significant role in devising quantum security in secret key generation and multiparty communication protocols.
In recent years, a significant amount of research has been devoted to understand the role of monogamy in various quantum phenomena, including violation of Bell inequalities and contextuality, and also in investigating correlation properties beyond quantum mechanics, such as in no-signalling theories. Lots of effort have also been spent in devising stronger monogamy conditions \cite{Eltschka09,Cornelio13,Kim14,Bai14a,Regula14} and to extend known features in discrete quantum states to continuous variable systems \cite{cv-mono}.

An important tool in quantitatively capturing the concept of monogamy of quantum correlations is the monogamy inequality, which bounds from above the distribution of quantum correlations among different parties in a multiparty state. The monogamy inequality can be studied in terms of the monogamy score. All monogamy inequality satisfying measures have a positive monogamy score, which is bounded above by 
a well defined value that can be derived as functions of the genuine multipartite entanglement of the quantum state. However, not all quantum correlation measures satisfy the monogamy inequality, and in general, the monogamy score can be negative. 
Interestingly, there exists no perceptible limiting value on the negative monogamy score. The main contribution of our work is the determination of a non-trivial lower bound of the monogamy score for measures and states where the monogamy inequality is violated. This is achieved using a complementarity relation between the normalized purity of a subsystem and the bipartite quantum correlation in the system. Subsequently, we analyze the strength and weakness of the lower bound for different quantum states and measures of quantum correlation, and observe conditions that immediately lead to monogamy. The results in the paper provide a unifying framework to study monogamy relations in both entanglement and information-theoretic quantum correlations, and are an important addition to the study of monogamy that opens possible directions for further investigation.

\begin{acknowledgments}
The authors would like to thank Ujjwal Sen for fruitful suggestions and discussions on the results. 
\end{acknowledgments}

\appendix
\section{\label{app} Quantum correlation measures and the complementarity relation}

\begin{figure}[]
\begin{center}
\subfloat[]{\includegraphics[width=4.1cm]{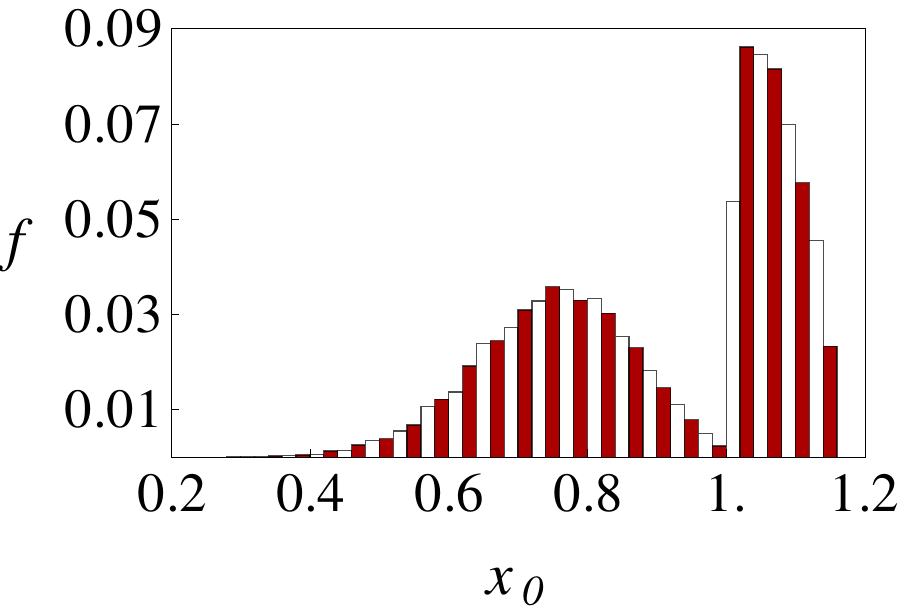}}\hspace{.2cm}
\subfloat[]{\includegraphics[width=4.1cm]{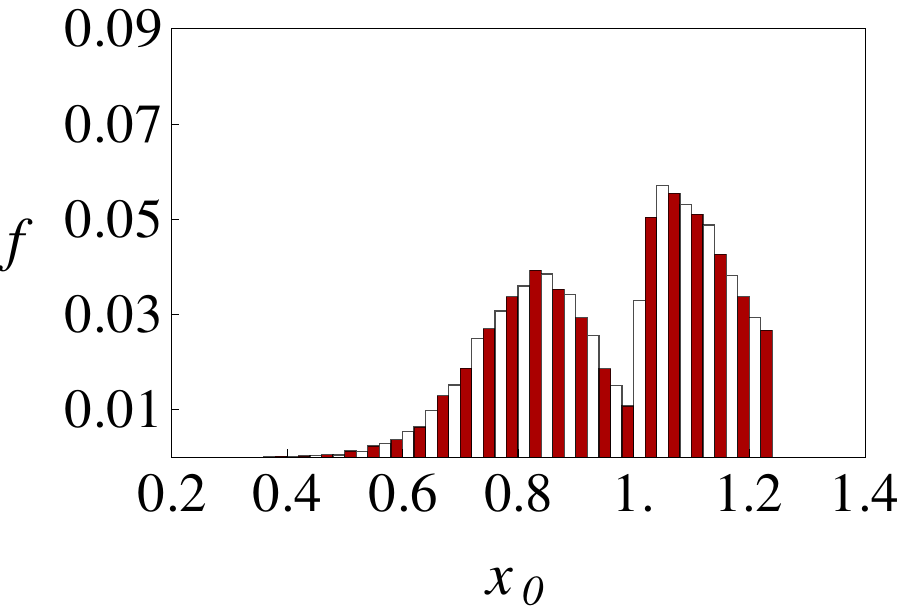}}\\
\subfloat[]{\includegraphics[width=4.1cm]{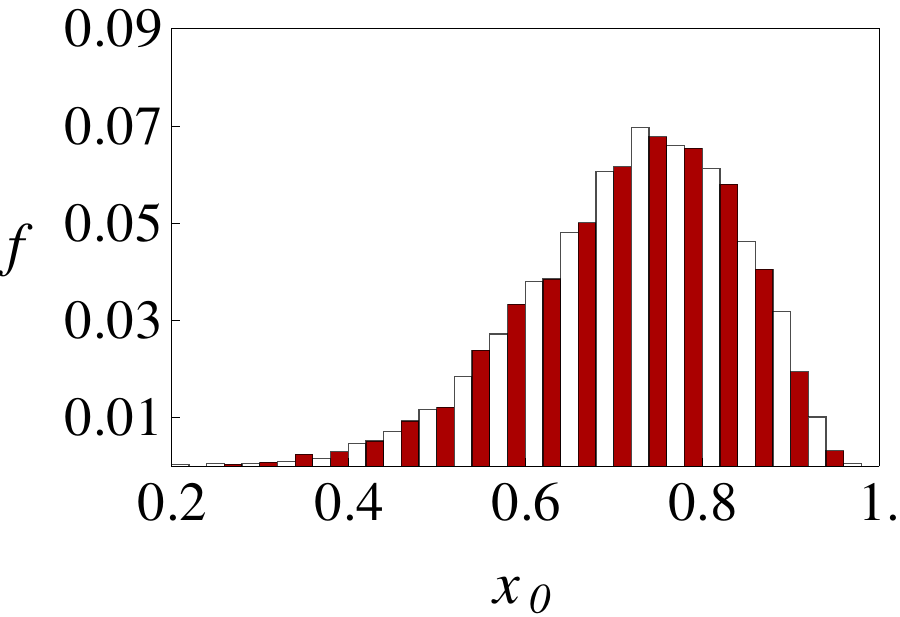}}\hspace{.2cm}
\subfloat[]{\includegraphics[width=4.1cm]{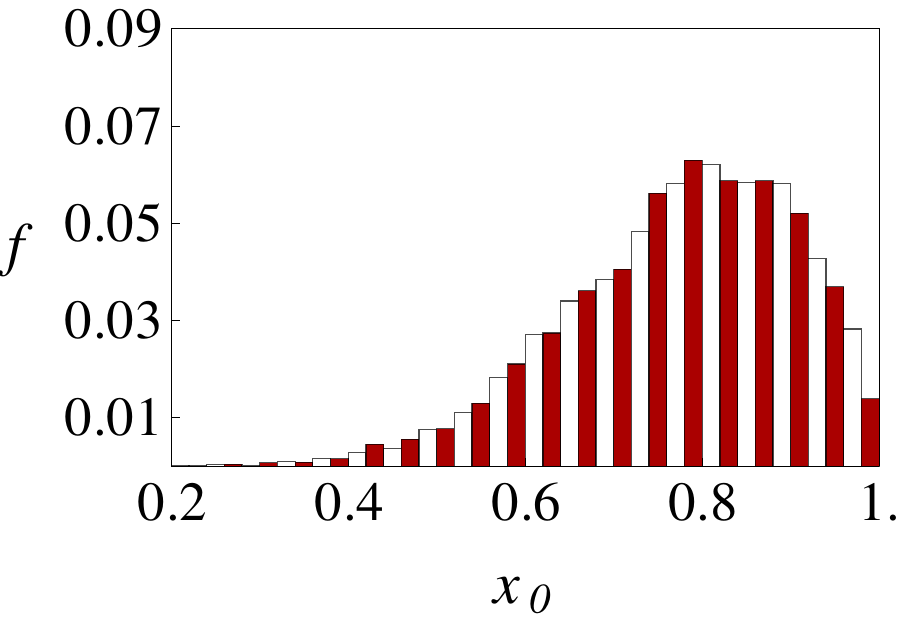}}
\caption{Histograms depicting the frequency ($f$) of states for different values of $x_0$, in the complementarity relation, $\mathbb{P}_A$ + $\mathbb{Q}_{A:BC} \leq x_0$. The graph corresponds to a set ($\sim 10^6$) of random Haar uniform generated rank-1 and rank-2 three-qubit states, and the quantum correlation measures, given by (a) negativity, (b) logarithmic negativity, (c) quantum discord, and (d) quantum work-deficit. For the measures $\mathcal{N}$ and $\mathcal{L}$, in subfigures (a) and (b), the rank-1 states correspond to the taller peaks at $x_0 \gtrsim$ 1, while the rank-2 states are evenly distributed in the region  $x_0 \lesssim$ 1. We note that for rank-1 states, $\mathbb{P}_A$ + $\mathbb{Q}_{A:BC}$ = $x_0$ = 1, for the measures $\mathcal{D}$ and $\Delta$. Hence, the graph for these measures, in subfigures (c) and (d), correspond only to rank-2 states.
We observe that $x_0$ is well below the trivial value, $x_0 <$ 2, for all states.}
\label{fig2}
\end{center}
\end{figure}

In our work, we investigate the lower bound on the monogamy score of quantum correlation measures that violate the monogamy inequality, using a complementarity relation between the normalized purity and normalized bipartite quantum correlation. In this Appendix, we discuss some entanglement, and information-theoretic quantum correlation measures that we have used in our study (see Figs.~\ref{threequbit} and \ref{fig2}), and look at the validity of the complementarity relation from the context of our results.


We begin with two measures of entanglement that are, in general, known to violate the monogamy inequality, namely, the \textit{negativity} \cite{negativity} and the \textit{logarithmic negativity} \cite{logneg}. 
Negativity (\({\cal N}\)) is a computable measure of bipartite entanglement, which is defined in terms of the eigenvalues of the partially transposed matrix \(\varrho_{AB}^{\Gamma_{A}}\). \({\cal N}\) is the sum of the absolute value of the negative eigenvalues of \(\varrho_{AB}^{\Gamma_{A}}\), or equivalently, ${\cal N}(\varrho_{AB})$ = $\frac{\|\varrho_{AB}^{\Gamma_A}\|_1-1}{2}$,
where $\|\cdot\|_1$ 
is the trace-norm.
For upto dimension $2 \times 3$, \({\cal N}\) = 0 for separable states. An extension of the entanglement measure of negativity, is the non-convex entanglement monotone, called the logarithmic negativity ($\mathcal{L}$). Mathematically, $\mathcal{L}$ = $\log_2\|\varrho_{AB}^{\Gamma_A}\|_1$, and in terms of $\mathcal{N}$, it is equal to $\log_2(2\mathcal{N} +1)$. Both $\mathcal{N}$ and $\mathcal{L}$ are non-monogamous entanglement monotones.

To analyze information-theoretic quantum correlations, we consider the measures of \textit{quantum discord} \cite{discord1} and \textit{quantum work-deficit} \cite{qwd}. Along the way, we also consider the \textit{quantum mutual information} \cite{qmi}, as a measure of total correlations, and the \textit{measured mutual information}, which has been characterized as classical part of total correlations \cite{discord1, modi}. Information-theoretic measures of quantum correlations draw upon specific properties of classical information theory \cite{thomas-cover}, and puts them under a quantum microscope. For example, consider two expressions of mutual information between two parties: $\mathcal{I}(\varrho_{AB})$ = $S(\varrho_A)$ + $S(\varrho_B)- S(\varrho_{AB})$, and
${\cal J}(\varrho_{AB})$ = $S(\varrho_B) - S(\varrho_{B|A})$. From a classical context, if $S(x)$ is the Shannon entropy and $\varrho_{AB}$, along with $\varrho_{A}$ and $\varrho_{B}$, are classical variables, the quantities $\mathcal{I}$ and $\mathcal{J}$ are equal. However, from a quantum perspective, where $S(x)$ is the von Neumann entropy and $\varrho_{AB}$ is a bipartite quantum density matrix, $\mathcal{I}$ is the quantum mutual information and $\mathcal{J}$ is the measured mutual information, and are, in general, not equal. $\mathcal{I}$, captures the total correlation between the two parties in $\varrho_{AB}$, whereas $\mathcal{J}$ captures the classical part of the  mutual information. This is due to the fact the conditional entropy, $S(\varrho_{B|A})$, used to define $\mathcal{J}$ involves local measurements on one of the subsytems, which washes away all quantum correlations. Quantum discord ($\cal{D}$), in a bipartite quantum state, is defined as the difference between the two quantities: $\cal{D}(\varrho_{AB})$ = $\cal{I}(\varrho_{AB})-\cal{J}(\varrho_{AB})$.
Another measure defined using an information perspective is the quantum work-deficit ($\Delta$), which is derived as the deficit between the amount of pure quantum states that are permitted for extraction under the set of closed global operations and product states that are obtained under a class of closed local operations and classical operations (closed LOCC). 
For a bipartite quantum state $\varrho_{AB}$, the number of pure qubits obtained under global operations, performed through sequential application of unitary and dephasing operations, is given by $\mathbb{I}_G (\varrho_{AB})$ = $\mathbb{N} - S(\rho_{AB})$, where $\mathbb{N}$ = $\log_2 d$. Here, $d$ is the dimension of the system. Under closed LOCC operations, optimized via local unitary and dephasing operations, and classical exchange of information, the number of pure classical states obtained is $\mathbb{I}_L(\varrho_{AB})$ = $\mathbb{N} - \inf_{\Lambda} [S(\varrho{'}_{AB})]$. $\Lambda$ belongs to the class of closed LOCC operations. The quantum correlation measure $\Delta$ is defined as: $\Delta$ = $\mathbb{I}_G - \mathbb{I}_L$.

The complementarity relation, defined in Eq.~(\ref{eq:cr}), holds for quantum correlation measures that satisfy the conditions, (i) ${\cal Q}(\varrho_{XY}) \leq S(\varrho_X)$, which holds for a host of quantum correlation measures, such as entanglement of formation, distillable entanglement, entanglement cost, quantum mutual information, relative entropy of entanglement, and classical correlation \cite{horodecki09, asu-comp},
(ii) $0 \leq {\cal Q}(\varrho_{XY}) \leq \log_2 d_Y$, which is known to be definitely satisfied by quantum mutual information. It should be noted that there may exist quantum correlation measures ${\mathbb Q}$ that may not satisfy one or both of the above conditions, or where the proof is not easily accessible, but may still satisfy the non-trivial complementarity relation given by,
$\mathbb{P}(\rho_{X}) + \mathbb{Q}(\rho_{XY}) \leq x_0$,
where $x_0 <$ 2. An analysis of the quantum correlation measures of negativity, logarithmic negativity, quantum discord, and quantum work-deficit, for random Haar uniform generated rank-1 and rank-2 three-qubit states are shown in Fig.~\ref{fig2}. A similar result was also observed in \cite{asu-comp}. We note that for rank-1 three-qubit states, $\mathbb{P}_A$ + $\mathbb{Q}_{A:BC}$ = 1, for $\mathcal{D}$ and $\Delta$. This is not the case for the entanglement measures, $\mathcal{N}$ and $\mathcal{L}$.


\begin{thebibliography}{99}
\bibitem{note1} In literature, purity of quantum state $\rho \in \mathbb{C}^d$ is often defined as $\mbox{Tr}(\rho^2)$. Here, however, we define purity as $\mathbb{P}(\rho)=({\log_2 d - S(\rho)})/{\log_2 d}$. For pure quantum states, both expressions yield unity.
\bibitem{comment} {For $x_0 \geq 1$, we have $1- x_0$ = $-t \leq$ 0, and from Eq.~(\ref{eq:lbound4a}), $\delta_{\mathbb{Q}} \geq -(n-1)(1 - \mathbb{P}_{A}) + t \geq -(n-1)(1 - \mathbb{P}_{A})$, since $t \geq 0$, which then gives us Eq.~(\ref{eq:lbound5}).}
\end{thebibliography}
\end{document}